\documentclass[%
print,
 amsmath,amssymb,
 aps,
]{revtex4}

\usepackage[sc]{mathpazo}
\usepackage{graphicx}
\usepackage{dcolumn}
\usepackage{bm}
\usepackage{dsfont}
\usepackage[landscape,papersize={297.1mm,210mm},left=1.9cm,right=1.6cm,top=2.7cm,bottom=2.8cm]{geometry}
\usepackage[                   
            pdfstartview=FitH,
            colorlinks, 
            pdfborder=001,   
            linkcolor=blue,
            anchorcolor=blue,
            citecolor=blue,
            urlcolor=blue
            ]{hyperref}
\usepackage{amsthm}
\makeatletter

\newcommand{\Rmnum}[1]{\expandafter\@slowromancap\romannumeral #1@}
\makeatother
\begin{document}
\title{Tighter uncertainty relations based on Wigner-Yanase skew information for observables and channels}
\author{Limei~Zhang,~Ting~Gao}
\email{gaoting@hebtu.edu.cn}
\affiliation {School of Mathematical Sciences,
 Hebei Normal University, Shijiazhuang 050024, China}
\author{Fengli~Yan}
\email{flyan@hebtu.edu.cn}
\affiliation {College of Physics, Hebei Normal University, Shijiazhuang 050024, China}

\begin{abstract}
Uncertainty principle is the basis of quantum mechanics. It reflects the basic law of the movement of microscopic particles. Wigner-Yanase skew information, as a measure of quantum uncertainties, is used to characterize the intrinsic features of the state and the observable. In this paper, we mainly investigate the sum uncertainty relations for both quantum mechanical observables and  quantum channels based on skew information. We establish a new uncertainty relation in terms of Wigner-Yanase skew information for $n$ observables, which is saturated (thus it holds as equality) for two incompatible observables. We also
present two uncertainty relations for arbitrary finite $N$ quantum channels by using skew information. Our uncertainty relations have tighter lower bounds than the existing ones. Detailed examples are provided.
\end{abstract}


\maketitle
\section{Introduction}
Uncertainty principle is one of the most essential features of quantum mechanics, and a fundamental difference between the quantum theory and classical theory. In 1927,  Heisenberg  first proposed the uncertainty relation of position-momentum \cite{Heisenberg1927}. Since the pioneering study, many groundbreaking results on the uncertainty relations were developed  \cite{Robertson1929,Schrdinger1930,MacconePRA2014}.

There exist many ways to express the uncertainty principle. The most famous one is Heisenberg-Robertson uncertainty relation \cite{Robertson1929} based on variance of measurement outcomes, which is usually stated: for any two observations $A$, $B$  and a quantum state $|\psi\rangle$, there is an uncertainty relation (UR)
\begin{equation}
\Delta A\Delta B\geq\frac{1}{2}\mid\langle\psi|[A,B]|\psi\rangle\mid,
\end{equation}
 where $\Delta M=\sqrt{\langle\psi|M^2|\psi\rangle-\langle\psi|M|\psi\rangle^2}$ and $[A,B]=AB-BA$. Subsequently, Schr$\ddot{o}$dinger obtained an improved UR \cite{Schrdinger1930},
\begin{equation}
\Delta A^2\Delta B^2\geq \mid\frac{1}{2}\langle[A,B]\rangle\mid^2+\mid\frac{1}{2}\langle\{A,B\}\rangle-\langle A\rangle\langle B\rangle\mid^2,
\end{equation}
where $\{A,B\}$ is the anticommutator  of $A$ and $B$, and $\langle M\rangle=\langle\psi|M|\psi\rangle$. Both these inequalities are state dependent and trivial when the state $|\psi\rangle$ of the system is an eigenstate of one observable; \textit{i.e.}, the lower bound can be  null whether the observables are compatible. Then Maccone \textit{et al.} \cite{MacconePRA2014} provided two stronger uncertainty relations based on the sum of variances
 \begin{equation}
 \begin{array}{rl}
 \Delta A^2+\Delta B^2\geq&\pm \texttt{i}\langle\psi|[A,B]|\psi\rangle+|\langle\psi|A\pm \texttt{i}B|\psi^{\perp}\rangle|^2,\\\\
\Delta A^2+\Delta B^2\geq&\frac{1}{2}\mid\langle\psi^{\perp}_{A+B}|A+B|\psi\rangle\mid^2.
\end{array}
\end{equation}
The first inequality is valid for arbitrary states $|\psi^{\perp}\rangle$ orthogonal to the state of the system $|\psi\rangle$, while  $|\psi^{\perp}_{A+B}\rangle\propto(A+B-\langle A+B\rangle)|\psi\rangle$ is a state orthogonal to $|\psi\rangle$.
These URs are guaranteed to be nontrivial whenever the two observables are incompatible on the state of the subsystem \cite{MacconePRA2014} and have been tested experimentally \cite{wangPRA2016}.
Since then some tighter URs based 
on variance were presented \cite{MondalPRA2017,SongSR2017,ZhangQIP2017}.

Entropy is another well-known way to express the uncertainty principle. Entropic uncertainty relation relating to any two observables was given by Deutsch \cite{Deutsch1983}, and later improved by Maassen and Uffink \cite{Maassen1988}, who obtained
\begin{equation}
  H(A)+H(B)\geq-2\log c(A,B),
\end{equation}
for incompatible observations $A$, $B$ with eigenbases $\{|a_i\rangle\}$, $\{|b_i\rangle\}$ respectively. Here  $c(A,B)=\max_{\{i,j\}}|\langle a_i|b_j\rangle|^2$, and $H(M)$ is the Shannon entropy of $M$. In addition, various URs relating to different entropies were proposed \cite{BirulaPRA2006,WilkPRA2009,TomamichelPRL2011}.  These entropic uncertainty relations are shown to be useful in various quantum information and computation tasks \cite{ColesRMP2017}.

Uncertainty relation, attracting sustained attentions, has many other ways to characterize \cite{Luo2000,Luo2003,RenesPRL2009,GrudkaPRA2013,Singh2016,YuanPRA2017,SazimPRA2018,SharmaPRA2018}. The skew information is one of them, that is, the skew information provides a new notion to quantify the Heisenberg uncertainty principle \cite{Luo2003}.  Here the skew information, introduced by Wigner-Yanase,  is denoted as \cite{skewinformation1963}
\begin{equation}\label{skew information defination}
 I_\rho(M)=-\frac{1}{2}\texttt{Tr}\left([\sqrt{\rho},M]^2\right)=\frac{1}{2}\parallel[\sqrt{\rho},M]\parallel^2.
\end{equation}
The Wigner-Yanase skew information characterizes the intrinsic features of the state $\rho$ and the observable $M$. Meanwhile, it satisfies all the expected requirements of an information measure \cite{skewinformation1963}. For pure states, skew information is the same as the variance \cite{LuoIEEE2004}, but there exist fundamental differences between them. For example, the skew information is convex \cite{LuoIEEE2004}, while the variance is concave; the skew information describes the  noncommutativity between the square root $\sqrt{\rho}$ and observable, while the variance describes the  noncommutativity between the state $\rho$ and the observable. The skew information can capture more quantum nature than the variance.

Recently, sum uncertainty relations based on  Wigner-Yanase skew information attracted attention. Chen \textit{et al.}  presented the sum uncertainty relation  relating to skew information for finite observables \cite{chenQIP2016}
\begin{equation}\label{chen bound}
    \sum_{i=1}^nI_{\rho}(M_i)\geq\frac{1}{n-2}\left[\sum_{1\leq i<j\leq n}I_{\rho}(M_i+M_j)-\frac{1}{(n-1)^2}\left(\sum_{1\leq i<j\leq n}\sqrt{I_{\rho}(M_i+M_j)}\right)^2\right],
\end{equation}
where $M_1,~M_2,\ldots,M_n $ are observables ( $n>2$ ), and $\rho$ is the state of the system. For convenience, the right hand of this inequality is labeled as $LB_0$. Obviously, $LB_0$ is always non-negative and vanishes only when $M_i$ are mutually commutative \cite{chenQIP2016}. Also another lower bound labeled as $LB_{\overline{0}}$ for $n$ observables \cite{chenQIP2016} was obtained,
 \begin{equation}\label{eigenvalue bound}
    \sum_{i=1}^nI_{\rho}(M_i)\geq\frac{1}{\lambda_{max}(G)}I_{\rho}(\sum_{i=1}^nM_i).
\end{equation}
Here $\lambda_{max}(G)$ is  the maximal eigenvalue of $G$, and $G$ is an $n$-dimensional Hermitian matrix with  entries $g_{ij}=\texttt{Tr}(X_iX_j)$, where $X_i=\frac{\texttt{i}[\sqrt{\rho},M_i]}{\parallel[\sqrt{\rho},M_i]\parallel}$, $\texttt{i}^2=-1$.

Quantum channel is very important in quantum theory. The URs for quantum channels have been widely studied \cite{ColesRMP2017,Krishna2002,MassarPRA2007}. More recently, Fu \textit{et al.} investigated the URs for quantum channels in terms of skew information \cite{FuQIP2019}. They  showed the uncertainty relation
\begin{equation}\label{Luo two channel uncertain}
 I_\rho(E_1)+ I_\rho(E_2)\geq\max_{\pi\in S_n}\frac{1}{2}\sum_{i=1}^nI_\rho(K^1_i\pm K_{\pi(i)}^2),
\end{equation}
for any two quantum channels $E_1$, $E_2$ with Kraus decompositions $E_1(\rho)=\sum\limits_{i=1}^nK^1_i\rho(K^1_i)^\dag$, $E_2(\rho)=\sum\limits_{i=1}^nK^2_i\rho(K^2_i)^\dag$ respectively, where  $\pi\in S_n$ is an arbitrary $n$-element permutation.

In this paper, we mainly investigate the sum uncertainty relations with Wigner-Yanase skew information. In Sec.\Rmnum{2}, we obtain a new tight UR  in terms of the skew information for arbitrary $n$  incompatible observables. The UR can be saturated  (thus it holds as equality)  for two observations. Then we compare our UR with previous work, and find that our lower bound is largest for any spin-$\frac{1}{2}$ particle. For  qutrit states, our UR is also tighter than the existing ones in \cite{chenQIP2016}. In Sec.\Rmnum{3}, we present two uncertainty relations for arbitrary finite $N$ quantum channels via the skew information, one of which is an equality when $N=2$ and better than the existing UR in \cite{FuQIP2019}. Finally, we conclude with a summary in Sec.\Rmnum{4}.
\section{Tighter Uncertainty relation for $n$ Hermitian operators}
In this section, we provide a tighter uncertainty relation via Wigner-Yanase skew information for arbitrary $n$ observables than the  existing ones. Before we present our main result, we first provide some \emph{Lemmas}, which are useful throughout this paper.

\emph{Lemma 1.} For any vectors $\textbf{a}_i$ in a Hilbert space, there is
\begin{equation}\label{Lemma 1}
\sum_{1\leq i<j\leq n}\parallel \textbf{a}_i+\textbf{a}_j\parallel^2=\parallel\sum_{i=1}^n\textbf{a}_i\parallel^2+(n-2)\sum_{i=1}^n\parallel \textbf{a}_i\parallel^2,
\end{equation}
where $\parallel\cdot\parallel$ is the norm of a vector.

\emph{Proof.}  It is easy to obtain
\begin{equation*}
\begin{array}{rl}
\sum\limits_{1\leq i<j\leq n}\parallel \textbf{a}_i+\textbf{a}_j\parallel^2=&(n-1)\sum\limits_i\parallel \textbf{a}_i\parallel^2+\sum\limits_{i\neq j}\langle \textbf{a}_i|\textbf{a}_j\rangle\\\\
=&\parallel\sum\limits_i\textbf{a}_i\parallel^2+(n-2)\sum\limits_{i}\parallel \textbf{a}_i\parallel^2,
\end{array}
\end{equation*}
as required. \qed

\emph{Lemma 2.} For any vectors $\textbf{a}_i$ of a Hilbert space, there is
\begin{equation}\label{Lemma 2}
\sum_{1\leq i<j\leq n}\| \textbf{a}_i-\textbf{a}_j\|^2=n\sum_i\|\textbf{a}_i\|^2-\|\sum_i\textbf{a}_i\|^2.
\end{equation}

\emph{Lemma 3.} For any vectors $\textbf{a}_i$ in a Hilbert space, there is an inequality
\begin{equation}\label{Lemma 3}
\frac{n(n-1)}{2}\sum_{1\leq i<j\leq n}\|\textbf{a}_i-\textbf{a}_j\|^2\geq\left(\sum_{1\leq i<j\leq n}\|\textbf{a}_i-\textbf{a}_j\|\right)^2.
\end{equation}

\emph{Proof.}  It follows immediately from the inequality~$(\sum\limits_{i=1}^nM_i)^2\leq n\sum\limits_{i=1}^nM_i^2$.\qed

Now, we are ready to have the following theorem, the tight sum uncertainty relation based on Wigner-Yanase skew information.

\emph{Theorem 1.} Let $M_1,~M_2,\cdots,M_n$ be arbitrary $n$ Hermitian operators. Then the sum uncertainty relation in terms of Wigner-Yanase skew information
\begin{equation}\label{tighter bound}
  \sum_{i=1}^nI_{\rho}(M_i)\geq\frac{1}{n}I_{\rho}(\sum_{i=1}^nM_i)+\frac{2}{n^2(n-1)}\left[\sum_{1\leq i<j\leq n}\sqrt{I_{\rho}(M_i-M_j)}\right]^2
\end{equation}
holds with equality when $n=2$. The lower bound is non-negative and  vanishes only when $M_i$ are mutually commutative.

\emph{Proof.}  For simplicity, the lower bound in $\eqref{tighter bound}$ is represented as $LB_1$. Obviously $LB_1$ is non-negative. If $LB_1$ is zero,  then $I_{\rho}(\sum_{i=1}^nM_i)=0$, $I_{\rho}(M_i-M_j)=0$ for $1\leq i<j\leq n$, which implies that $I_{\rho}(M_i)=0$, $i=1,2,\cdots,n$. Hence we get $\langle[M_i,M_j]\rangle_{\rho}=0$.

Due to the Lemma 2 and  Lemma 3, one has
\begin{equation*}\begin{array}{rl}
  \frac{2}{n^2(n-1)}\left(\sum\limits_{1\leq i<j\leq n}\|\textbf{a}_i-\textbf{a}_j\|\right)^2\leq&\frac{1}{n}\sum\limits_{1\leq i<j\leq n}\|\textbf{a}_i-\textbf{a}_j\|^2\\\\
  =&\left(\sum\limits_i\|\textbf{a}_i\|^2-\frac{1}{n}\|\sum\limits_i\textbf{a}_i\|^2\right),
\end{array}
\end{equation*}
that is
\begin{equation}\label{inequality}
  \sum_{i=1}^n\|\textbf{a}_i\|^2\geq \frac{1}{n}\| \sum_{i=1}^n \textbf{a}_i\|^2+\frac{2}{n^2(n-1)}\left(\sum_{1\leq i<j\leq n}\|\textbf{a}_i-\textbf{a}_j\|\right)^2.
\end{equation}
One then obtains inequality $\eqref{tighter bound}$ by substituting $\textbf{a}_i$ with $[\sqrt{\rho},M_i]$.

Specially, for two incompatible observables $M_1$ and $M_2$, using equality $\|\textbf{a}+\textbf{b}\|^2+\|\textbf{a}-\textbf{b}\|^2=2(\|\textbf{a}\|^2+\|\textbf{b}\|^2)$, we have
\begin{equation}\label{Corollary}
  I_{\rho}(M_1)+I_{\rho}(M_2)=\frac{1}{2}[I_{\rho}(M_1+M_2)+I_{\rho}(M_1-M_2)].
  \end{equation}
Hence UR $\eqref{tighter bound}$ is saturated for two observations. \qed

Clearly, when $n=2$, our formula in \emph{Theorem 1} gives a  tighter uncertainty relation than the result $I_{\rho}(M_1)+I_{\rho}(M_2)\geq \frac{1}{2}\max{\{I_{\rho}(M_1+M_2),~I_{\rho}(M_1-M_2)\}}$ in \cite{chenQIP2016}.

\emph{Corollary.} For a spin-$\frac{1}{2}$ particle, when Pauli-spin operators $\sigma_x,~\sigma_y,~\sigma_z$ are chosen as the observables, the lower bound $LB_1$, the right hand of $\eqref{tighter bound}$, is tighter than $LB_0$.

\emph{Proof.} Given a qubit state $\rho$
\begin{equation}\label{operation definition}
  \rho=\frac{1}{2}(\mathcal{I}+\vec{r}\cdot\vec{\sigma}),
\end{equation}
where $\vec{r}=(x,~y,~z)$ is a real three-dimensional vector such that $\|\vec{r}\|\leq1$, and $\vec{\sigma}=(\sigma_x,~\sigma_y,~\sigma_z)$ is constituted by Pauli matrices. Then the density operator $\rho$ has a spectral decomposition $\rho=\sum_i\lambda_i|\psi_i\rangle\langle\psi_i|$, where $\lambda_1=\frac{1-\sqrt{t}}{2},~\lambda_2=\frac{1+\sqrt{t}}{2}$, $|\psi_1\rangle=\frac{1}{\sqrt{2t-2z\sqrt{t}}}\left[(z-\sqrt{t})|0\rangle+(x+y\texttt{i})|1\rangle\right]$, and $|\psi_2\rangle=\frac{1}{\sqrt{2t+2z\sqrt{t}}}\left[(z+\sqrt{t})|0\rangle+(x+y\texttt{i})|1\rangle\right]$,  with $t=\|\vec{r}\|^2=x^2+y^2+z^2$.

For Pauli operators, the sum of skew information is
\begin{equation}\label{xinxihe}
I_{\rho}(\sigma_x)+ I_{\rho}(\sigma_y)+ I_{\rho}(\sigma_z)=2\left(1-\sqrt{1-t}\right).
\end{equation}

From the UR (\ref{tighter bound}) we have the lower bound of (\ref{xinxihe})
\begin{equation}\label{qubit tighter bound 2}
\begin{array}{rl}
 I_{\rho}(\sigma_x)+ I_{\rho}(\sigma_y)+ I_{\rho}(\sigma_z)\geq &\frac{2}{3}\Big(1-\sqrt{1-t}\Big)\left(1-\frac{xy+xz+yz}{t}\right)+\frac{1}{9}\alpha^2,
 \end{array}
\end{equation}
while the UR (\ref{chen bound}) gives
\begin{equation}\label{chen bound 2}
\begin{array}{rl}
 I_{\rho}(\sigma_x)+ I_{\rho}(\sigma_y)+ I_{\rho}(\sigma_z)\geq &\Big(1-\sqrt{1-t}\Big)\left[4-\frac{2\big(xy+xz+yz\big)}{t}\right]-\frac{1}{4}\beta^2,
 \end{array}\end{equation}
where
\begin{equation}\begin{array}{rl}
\alpha=&\sqrt{1-\sqrt{1-t}}\left(\sqrt{1+\frac{z^2+2xy}{t}}+\sqrt{1+\frac{y^2+2xz}{t}}+\sqrt{1+\frac{x^2+2yz}{t}}\right)\\\\
\beta=&\sqrt{1-\sqrt{1-t}}\left(\sqrt{1+\frac{z^2-2xy}{t}}+\sqrt{1+\frac{y^2-2xz}{t}}+\sqrt{1+\frac{x^2-2yz}{t}}\right).
\end{array}\end{equation}

Comparing our lower bound $LB_1$ \big(the right hand of inequality (\ref{qubit tighter bound 2})\big) with the lower bound $LB_0$ \big(the right hand of inequality (\ref{chen bound 2})\big), we obtain the difference value
 \begin{equation}
LB_1-LB_0=\left(1-\sqrt{1-t}\right)\gamma,
\end{equation}
where
\begin{equation}\label{cha zhi}
\begin{array}{rl}
\gamma=&-\frac{10}{3}+\frac{4\big(xy+xz+yz\big)}{3t}+\frac{1}{9}\left(\sqrt{1+\frac{z^2+2xy}{t}}+\sqrt{1+\frac{y^2+2xz}{t}}+\sqrt{1+\frac{x^2+2yz}{t}}\right)^2\\\\
&+\frac{1}{4}\left(\sqrt{1+\frac{z^2-2xy}{t}}+\sqrt{1+\frac{y^2-2xz}{t}}+\sqrt{1+\frac{x^2-2yz}{t}}\right)^2.
\end{array}
\end{equation}
Note that $1-\sqrt{1-t}\geq0$, hence we need only consider the  algebraic expression $\gamma$.

Let $x=\sqrt{t}\sin\theta\cos\varphi,~y=\sqrt{t}\sin\theta\sin\varphi,~z=\sqrt{t}\cos\theta$, $\theta\in[0,~\pi],~\varphi\in[0,~2\pi]$, then we derive
 \begin{equation}
\begin{array}{rl}
\gamma=&-\frac{10}{3}+\frac{4}{3}\left(\sin^2\theta\cos\varphi\sin\varphi+\sin\theta\cos\theta\cos\varphi+\sin\theta\cos\theta\sin\varphi\right)\\\\
&+\frac{1}{9}\Big(\sqrt{1+\cos^2\theta+2\sin^2\theta\cos\varphi\sin\varphi}+\sqrt{1+\sin^2\theta\sin^2\varphi+2\sin\theta\cos\theta\cos\varphi}\\\\
&+\sqrt{1+\sin^2\theta\cos^2\varphi+2\sin\theta\cos\theta\sin\varphi}~\Big)^2+\frac{1}{4}\Big(\sqrt{1+\cos^2\theta-2\sin^2\theta\cos\varphi\sin\varphi}\\\\
&+\sqrt{1+\sin^2\theta\sin^2\varphi-2\sin\theta\cos\theta\cos\varphi}+\sqrt{1+\sin^2\theta\cos^2\varphi-2\sin\theta\cos\theta\sin\varphi}~\Big)^2\\\\
\geq&\sqrt{3}-\frac{4}{3}\approx0.398,
\end{array}
\end{equation}
as depicted in Fig.1.
\begin{center}\label{Fig0}
\includegraphics[width=8cm]{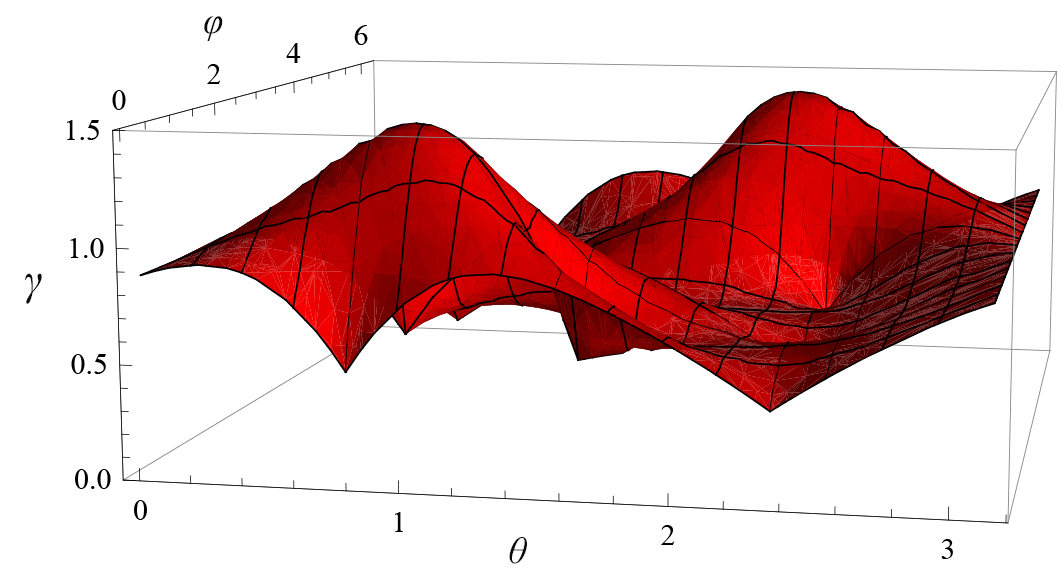}\\
\textbf{Fig.1}~~Function $\gamma$
\end{center}

Therefore, our lower bound $LB_1$  is strictly larger than the lower bound $LB_0$ except $t=0$ for a spin-$\frac{1}{2}$  particle. \qed

\emph{Example 1.} Suppose $\rho$ is a qubit state with Bloch vector $\vec{r}=(\frac{\sqrt{3}}{2}\cos\theta,\frac{\sqrt{3}}{2}\sin\theta,0)$, and Pauli operators $\sigma_x,~\sigma_y,~\sigma_z$ are three observables. Then our lower bound $LB_1=\frac{1}{6}(2-\sin2\theta)+\frac{1}{36}\left(\sqrt{2+2\sin2\theta}+\sqrt{3-\cos2\theta}+\sqrt{3+\cos2\theta}\right)^2$,
 the lower bound in $\eqref{chen bound}$ $LB_0=2-\frac{1}{2}\sin2\theta-\frac{1}{16}\left(\sqrt{2-2\sin2\theta}+\sqrt{3-\cos2\theta}+\sqrt{3+\cos2\theta}\right)^2$ and lower bound in $\eqref{eigenvalue bound}$ $LB_{\overline{0}}=\frac{1}{4}(2-\sin2\theta)$.
 These lower bounds  are illustrated in Fig.2. One can see that our result is much better than the results in \cite{chenQIP2016}.
\begin{center}\label{Fig1}
\includegraphics[width=8cm]{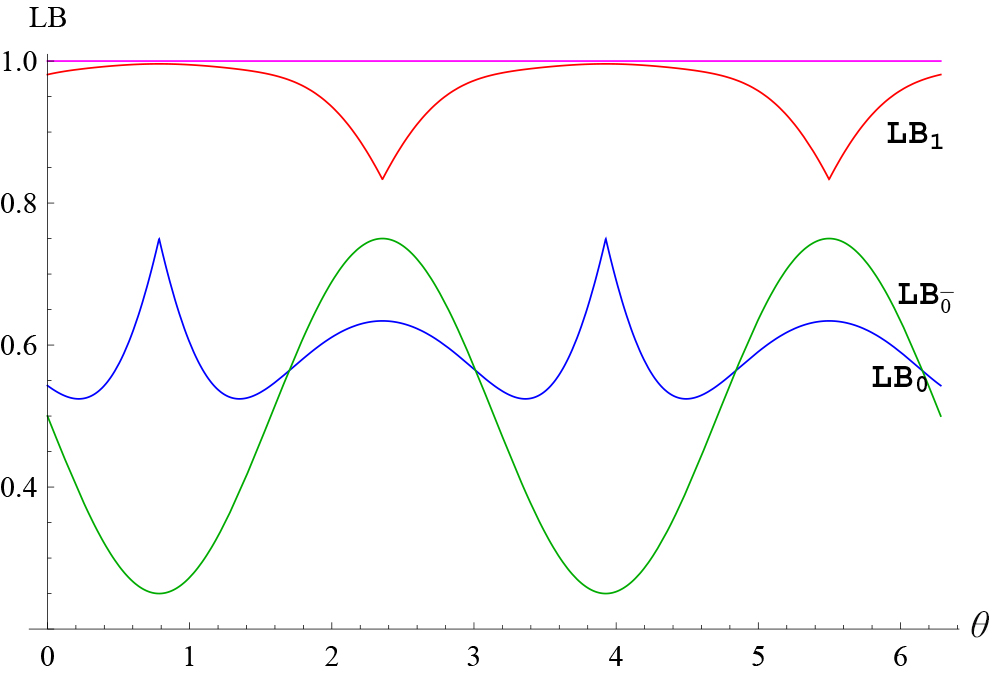}
\end{center}
{\centering\textbf{Fig.2}}~~Comparison of  the lower bound $LB_1$, lower bound $LB_0$ and lower bound $LB_{\overline{0}}$ for the qubit state $\rho$ with Bloch vector $\vec{r}=(\frac{\sqrt{3}}{2}\cos\theta,\frac{\sqrt{3}}{2}\sin\theta,0)$.
 The magenta line represents the sum $I_{\rho}(\sigma_x)+ I_{\rho}(\sigma_y)+ I_{\rho}(\sigma_z)$, the red line represents our lower bound $LB_1$, blue line represents lower bound $LB_0$, and the green line represents lower bound $LB_{\overline{0}}$. Obviously, our lower bound is the tightest one.

\emph{Example 2.} We consider a spin-1 particle in the state
\begin{equation}\label{spin-1 state}
  |\psi\rangle=\sin\theta\cos\varphi|1\rangle+\sin\theta\sin\varphi|0\rangle+\cos\theta|-1\rangle,
\end{equation}
with $0\leq\theta\leq\pi,~0\leq\varphi\leq2\pi$, and choose the angular momentum operators \\
\begin{center}
$L_x=\frac{1}{\sqrt{2}}\left(
  \begin{array}{ccc}
    0 & 1 & 0 \\
    1 & 0 & 1 \\
    0 & 1 & 0 \\
  \end{array}
\right)$,~~~~~~$L_y=\frac{1}{\sqrt{2}}\left(
  \begin{array}{ccc}
    0 & -\texttt{i} & 0 \\
    \texttt{i} & 0 & -\texttt{i} \\
    0 & \texttt{i} & 0 \\
  \end{array}
\right)$,~~~~~$L_z=\left(
  \begin{array}{ccc}
    1 & 0 & 0 \\
    0 & 0 & 0 \\
    0 & 0 & -1 \\
  \end{array}\right)$
\end{center} as observables.

For angular momentum operators and quantum state $|\psi\rangle$, the sum of skew information is
 \begin{equation}
 I_{\rho}(L_x)+I_{\rho}(L_y)+I_{\rho}(L_z)=2-2\sin^2\theta\sin^2\varphi\left(\cos\theta+\sin\theta\cos\varphi\right)^2-\big(\cos^2\theta-\sin^2\theta\cos^2\varphi\big)^2.
 \end{equation}
 The lower bound $LB_1$ of $I_{\rho}(L_x)+ I_{\rho}(L_y)+ I_{\rho}(L_z)$ in $\eqref{tighter bound}$ and the lower bound $LB_0$ in $\eqref{chen bound}$
are illustrated in Fig.3. Our lower bound is tighter than lower bound $LB_0$  for this state $|\psi\rangle$.
\begin{flushleft}\label{Fig2}
\includegraphics[width=7.9cm]{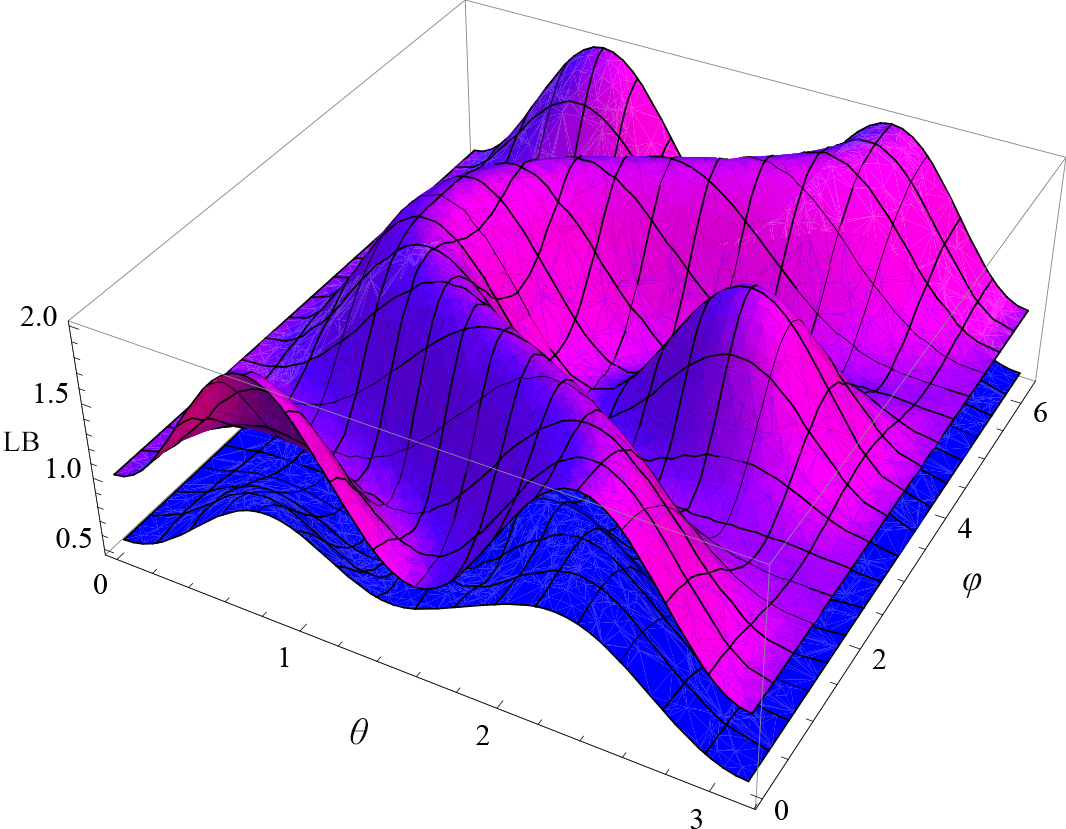}~~~~~\includegraphics[width=7.9cm]{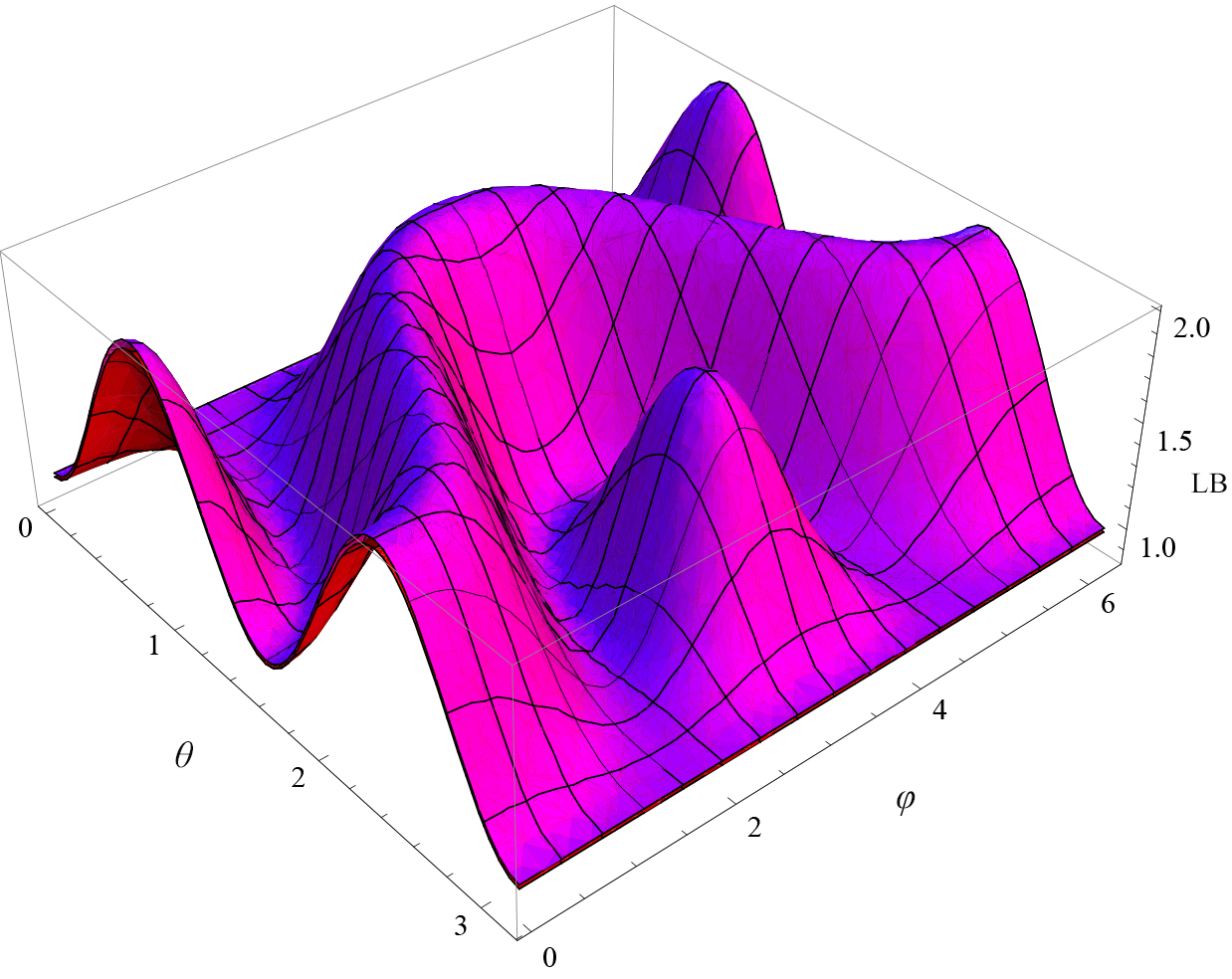}\\
~~~~~~~~~~~~~~~~~~~~~~~~~~~~~~~~~(a)~~~~~~~~~~~~~~~~~~~~~~~~~~~~~~~~~~~~~~~~~~~~~~~~~~~~~~~~~~~~~~~~~~~~~~~~~~~~~~~(b)
\end{flushleft}
\begin{flushleft}
 \includegraphics[width=8cm]{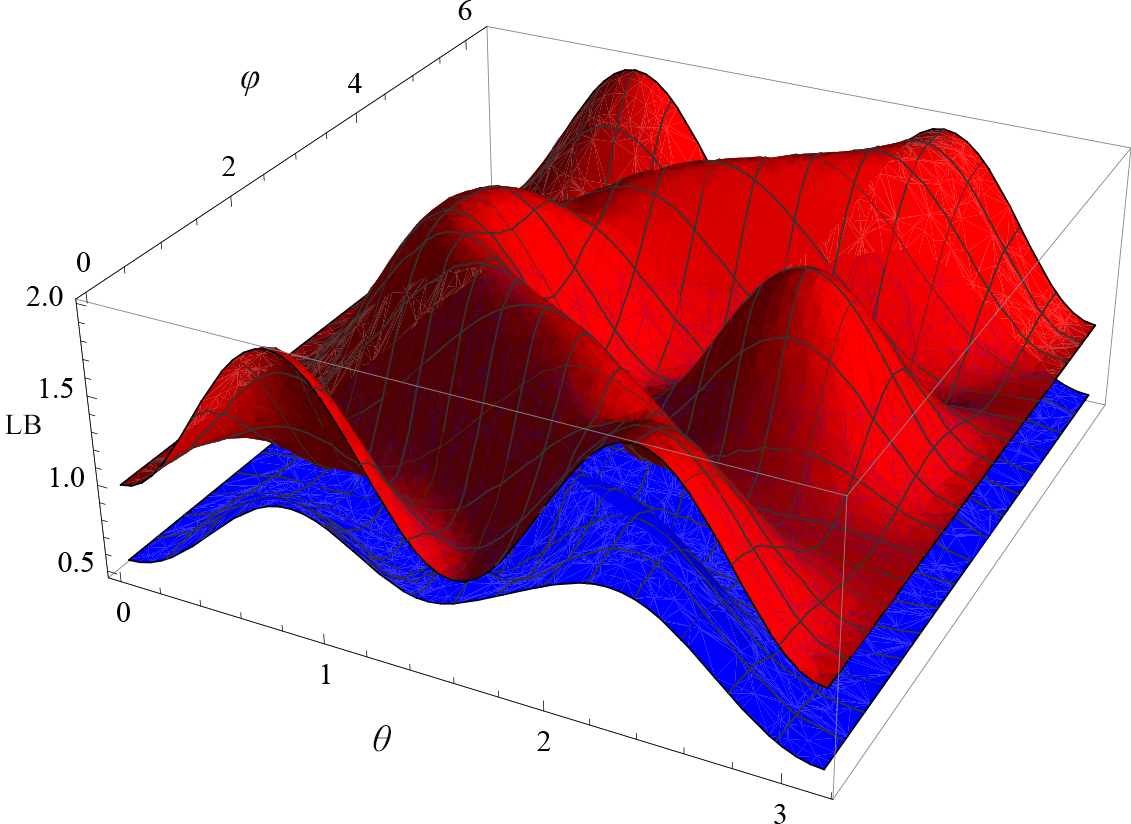}~~~~~~~~~\includegraphics[width=8cm]{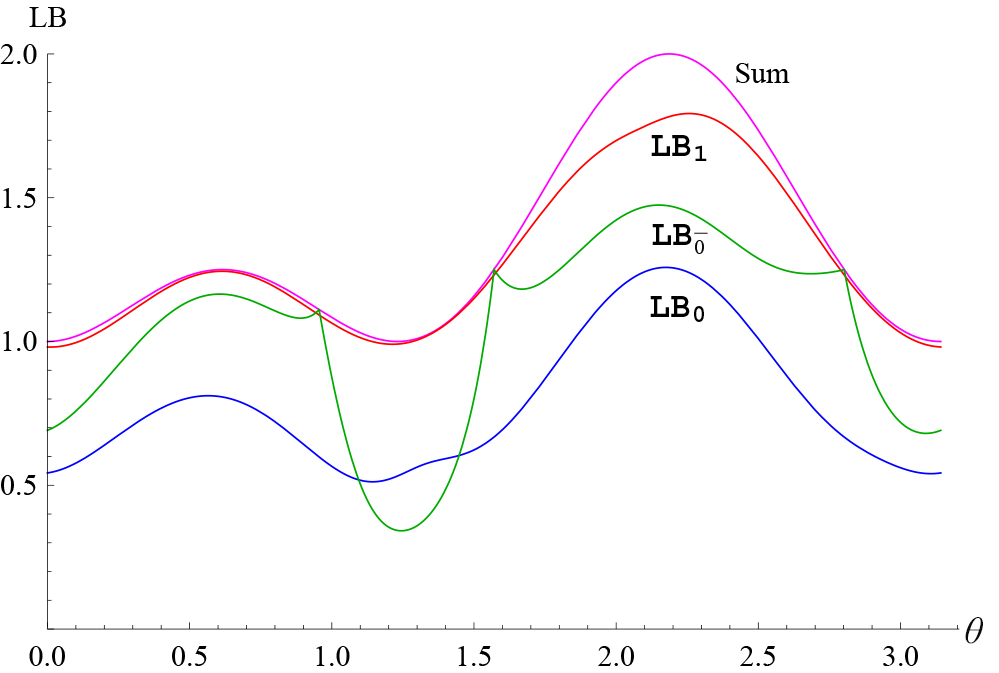}\\
~~~~~~~~~~~~~~~~~~~~~~~~~~~~~~~~~(c)~~~~~~~~~~~~~~~~~~~~~~~~~~~~~~~~~~~~~~~~~~~~~~~~~~~~~~~~~~~~~~~~~~~~~~~~~~~~~~~~~~~~~~(d)
\end{flushleft}

{\centering\textbf{Fig.3}}~~The comparison of different lower bounds referring  hereinbefore for $\rho=|\psi\rangle\langle\psi|$ with $|\psi\rangle$ in $\eqref{spin-1 state}$ and observables $L_x,~L_y,~L_z$. The magenta surface (line), the red surface  (line), the green surface  (line), and the blue surface  (line) represent the sum of the skew information, our lower bound  $LB_1$,  the lower bound $LB_{\overline{0}}$, and the lower bound $LB_0$, respectively.
(a) The comparison of lower bound $LB_0$ with the sum of the skew information; (b)   The comparison of our bound $LB_1$ with the sum of the skew information; (c) The comparison of  $LB_1$  with $LB_0$; (d) The comparison of the sum of skew information $I_{\rho}(L_x)+ I_{\rho}(L_y)+ I_{\rho}(L_z)$, the lower bounds $LB_0$, $LB_{\overline{0}}$ ~and $LB_1$ for $\varphi=\frac{\pi}{4}$.

\emph{Example 3.} For a qutrit state
\begin{center}\label{liangzitai}
$\rho=\frac{1}{3}\left(
  \begin{array}{ccc}
    1 & -\sqrt{3}a\texttt{i}\cos\alpha & -\sqrt{3}a\texttt{i}\sin\alpha\cos\beta \\
    \sqrt{3}a\texttt{i}\cos\alpha & 1 & -\sqrt{3}a\texttt{i}\sin\alpha\sin\beta  \\
    \sqrt{3}a\texttt{i}\sin\alpha\cos\beta & \sqrt{3}a\texttt{i}\sin\alpha\sin\beta  & 1 \\
  \end{array}
\right)$
\end{center}
with $|a|\leq\frac{1}{\sqrt{3}}$, $0<\alpha<\pi$, and $0\leq\beta\leq2\pi$, we choose the angular momentum operators \\
\begin{center}
$L_x=\frac{1}{\sqrt{2}}\left(
  \begin{array}{ccc}
    0 & 1 & 0 \\
    1 & 0 & 1 \\
    0 & 1 & 0 \\
  \end{array}
\right)$,~~~~~~$L_y=\frac{1}{\sqrt{2}}\left(
  \begin{array}{ccc}
    0 & -\texttt{i} & 0 \\
    \texttt{i} & 0 & -\texttt{i} \\
    0 & \texttt{i} & 0 \\
  \end{array}
\right)$,~~~~~$L_z=\left(
  \begin{array}{ccc}
    1 & 0 & 0 \\
    0 & 0 & 0 \\
    0 & 0 & -1 \\
  \end{array}\right)$
\end{center} as observables.

For angular momentum operators and quantum state $\rho$, if $|a|=\frac{1}{\sqrt{3}}$, the sum of skew information $I_{\rho}(L_x)+ I_{\rho}(L_y)+ I_{\rho}(L_z)$, the lower bound $LB_1$ in $\eqref{tighter bound}$ and the lower bound $LB_0$ in $\eqref{chen bound}$ are illustrated in Fig.4. It is worth pointing out that our lower bound is tighter than lower bound $LB_0$  for this state $\rho$.

\begin{flushleft}\label{Fig3}
\includegraphics[width=8cm]{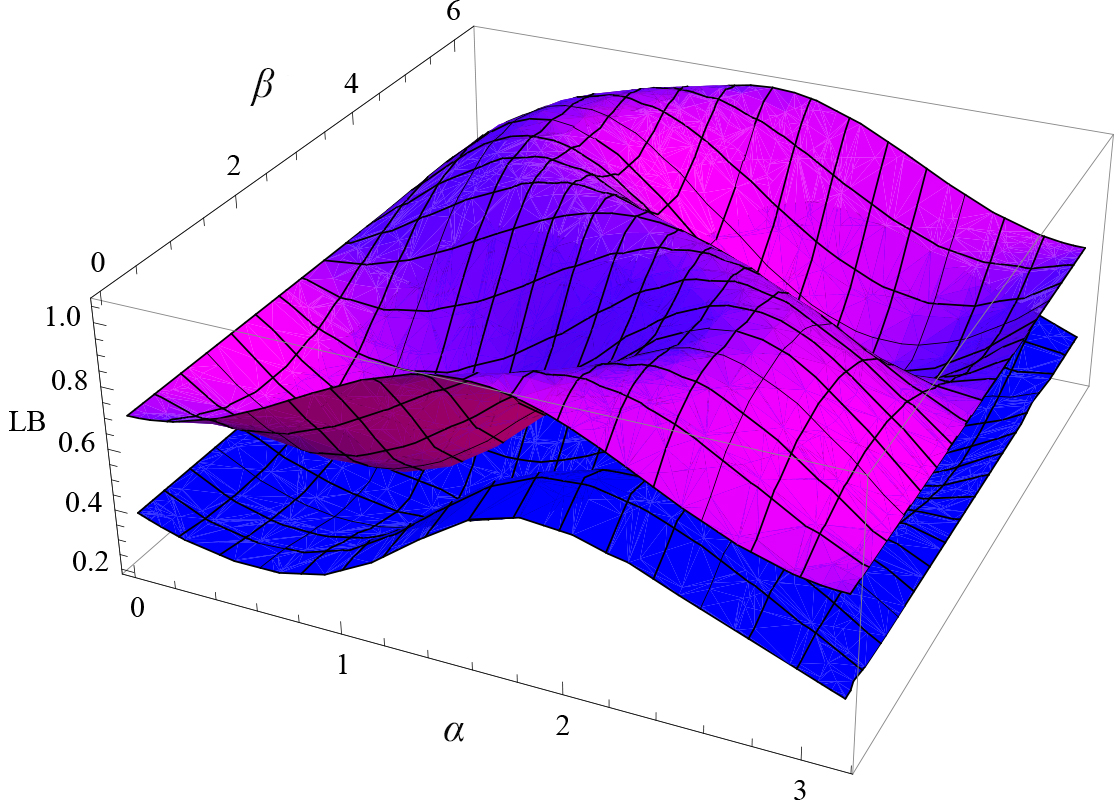}~~~~~\includegraphics[width=8cm]{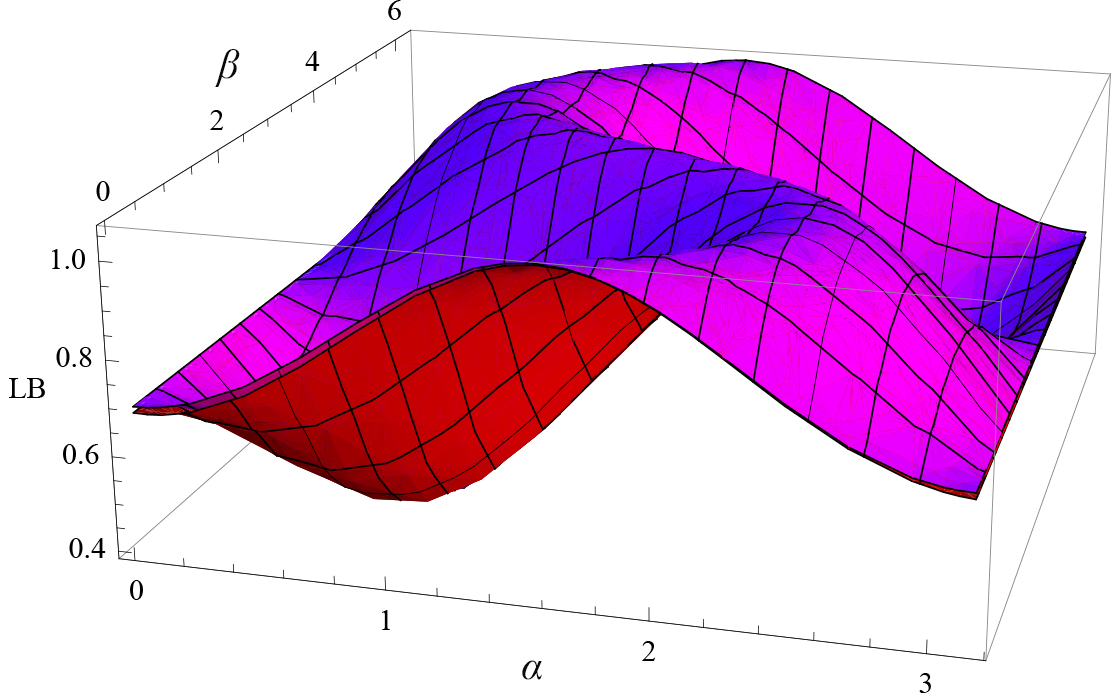}\\
~~~~~~~~~~~~~~~~~~~~~~~~~~~~~~~~~~~~~~~~~~(e)~~~~~~~~~~~~~~~~~~~~~~~~~~~~~~~~~~~~~~~~~~~~~~~~~~~~~~~~~~~~~~~~~~~~~~(f)
\end{flushleft}
\begin{center}
\includegraphics[width=8cm]{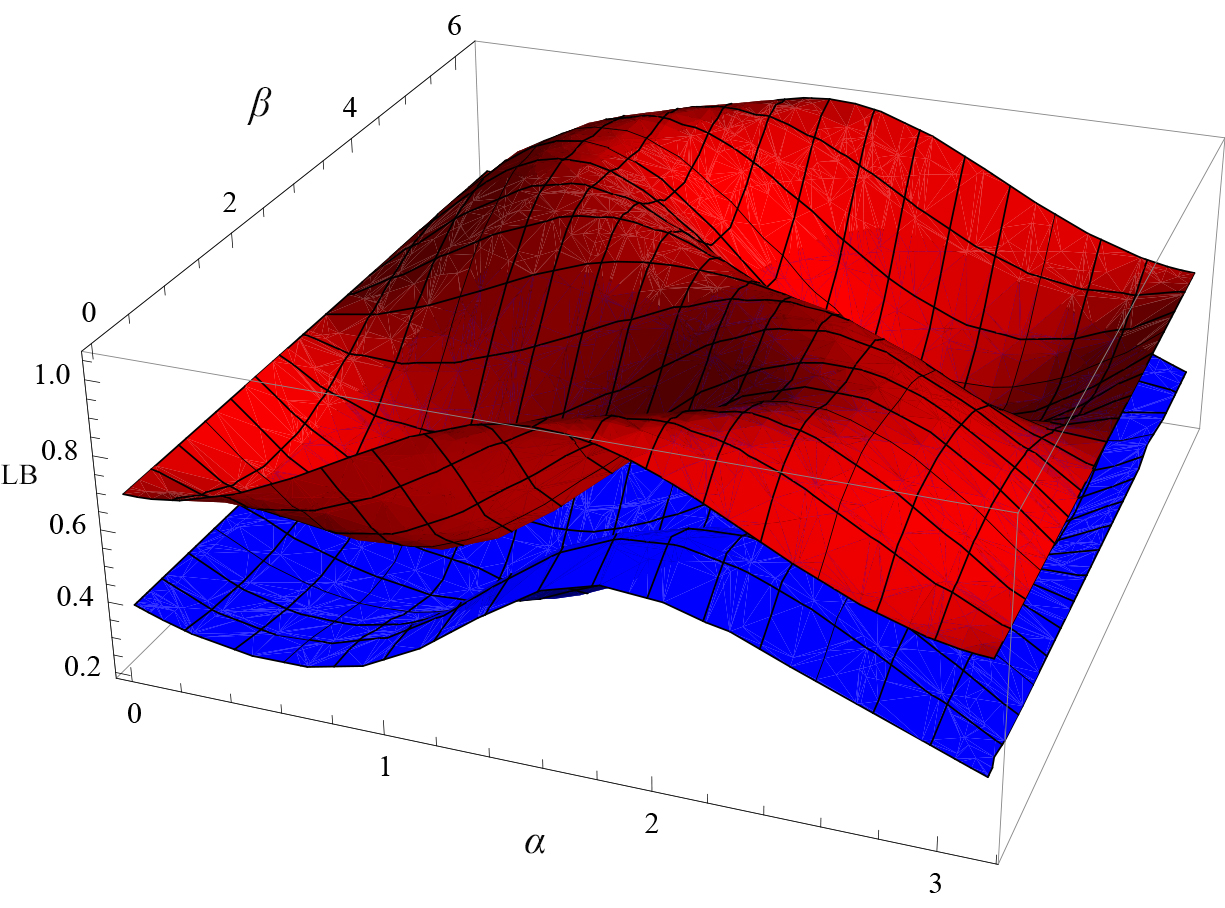}~~~~~\includegraphics[width=8cm]{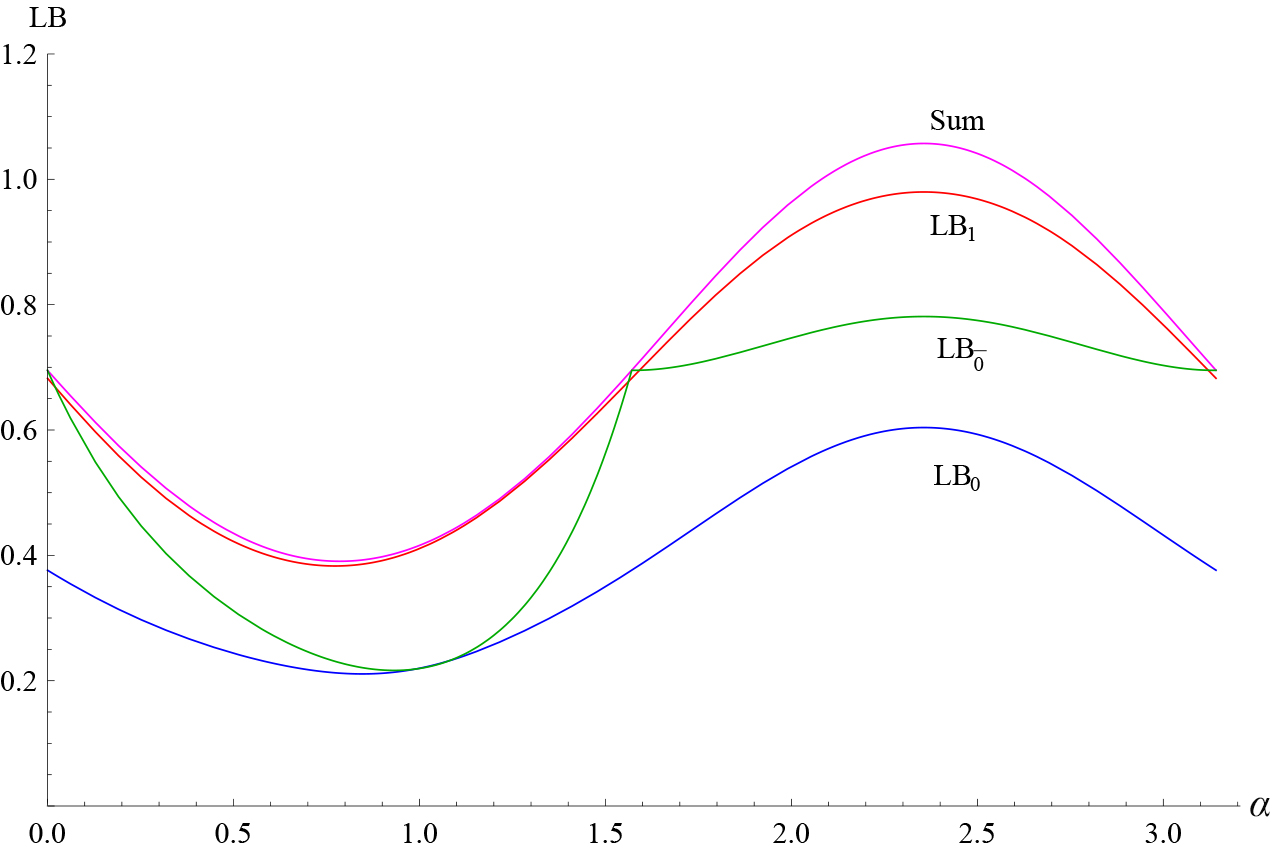}\\
(g)~~~~~~~~~~~~~~~~~~~~~~~~~~~~~~~~~~~~~~~~~~~~~~~~~~~~~~~~~~~~~~~~~~~~~~~(h)
\end{center}
{\centering\textbf{Fig.4}}~~The comparison of different lower bounds for $\rho$ with $|a|=\frac{1}{\sqrt{3}}$ and observables $L_x,~L_y,~L_z$. The magenta surface (line), the red surface  (line), the green surface  (line), and the blue surface  (line) represent the sum of the skew information, our lower bound  $LB_1$,  the lower bound $LB_{\overline{0}}$, and the lower bound $LB_0$, respectively.
(e) The comparison of lower bound $LB_0$ with the sum of the skew information; (f)   The comparison of our bound $LB_1$ with the sum of the skew information; (g) The comparison of  $LB_1$  with $LB_0$;
(h) The comparison of the sum of skew information $I_{\rho}(L_x)+ I_{\rho}(L_y)+ I_{\rho}(L_z)$, the lower bounds $LB_0$, $LB_{\overline{0}}$ ~and $LB_1$ for $\beta=\frac{\pi}{2}$. Our lower bound is larger than the lower bound in \cite{chenQIP2016}.
\section{Uncertainty relations for quantum channels}
The Wigner-Yanase skew information $I_\rho(A)$ is  initially defined for a Hermitian operator $A$, whereas recently it has been generalized to arbitrary operators \cite{FanQIP2018,LuoPRA2018,skewinformation2019}. Let $E$ be a quantum channel with Kraus representation $E(\rho)=\sum\limits_{i=1}^nK_i\rho(K_i)^\dag$, where $K_i$ are not necessarily Hermitian operators. The skew information of $\rho$ with respect to the channel is
\begin{equation}\label{channel skew information}
  I_\rho(E)=\sum_{i=1}^nI_\rho(K_i),
\end{equation}
where $I_\rho(K_i)=\frac{1}{2}\texttt{Tr}[\sqrt{\rho},K_i]^\dag[\sqrt{\rho},K_i]$ \cite{LuoPRA2018,skewinformation2019}. $ I_\rho(E)$ is related with the state $\rho$ and channel, which reveals some intrinsic features of the state-channel interaction \cite{skewinformation2019,FuQIP2019}. In \cite{FuQIP2019}, Fu \textit{et al.} gave uncertainty relation $\eqref{Luo two channel uncertain}$ for two quantum channels. In the following, we present uncertainty relations for arbitrary $N$ quantum channels. Our UR is tighter than $\eqref{Luo two channel uncertain}$.

\emph{Theorem 3.} Let $E_1,~E_2,\cdots,E_N$ be arbitrary $N$ ($N>2$) channels, and  each channel has Kraus representation $E_t(\rho)=\sum\limits_{i=1}^nK^t_i\rho(K^t_i)^\dag$, $t=1, 2,\cdots, N$. One has
\begin{equation}\label{N channel uncertain 1}
   \sum_{t=1}^NI_{\rho}(E_t)\geq\max_{\pi_t,\pi_s\in S_n}\frac{1}{N-2}\left\{\sum_{1\leq t<s\leq N}\sum_{i=1}^nI_{\rho}\left(K_{\pi_t(i)}^t+K_{\pi_s(i)}^s\right)-\frac{1}{(N-1)^2}\left[\sum_{i=1}^n\left(\sum_{1\leq t<s\leq N}\sqrt{I_{\rho}\left(K_{\pi_t(i)}^t+K_{\pi_s(i)}^s\right)}\right)^2\right]\right\}
\end{equation}
where $\pi_t\in S_n$ is an arbitrary $n$-element permutation.

\emph{Proof.}  By the inequality in \cite{SRChen}
\begin{equation*}
   \sum_{t=1}^N\|\textbf{a}_t\|^2\geq\frac{1}{N-2}\left[\sum_{1\leq t<s\leq N}\|\textbf{a}_t+\textbf{a}_s\|^2-\frac{1}{(N-1)^2}\left(\sum_{1\leq t<s\leq N}(\|\textbf{a}_t+\textbf{a}_s\|)\right)^2\right],
\end{equation*}
one has
\begin{equation}
   \sum_{t=1}^NI_{\rho}(K_{\pi_t(i)}^t)\geq\frac{1}{N-2}\left[\sum_{1\leq t<s\leq N}I_{\rho}\left(K_{\pi_t(i)}^t+K_{\pi_s(i)}^s\right)-\frac{1}{(N-1)^2}\left(\sum_{1\leq t<s\leq N}\sqrt{I_{\rho}\left(K_{\pi_t(i)}^t+K_{\pi_s(i)}^s\right)}\right)^2\right],
\end{equation}
which implies inequality $\eqref{N channel uncertain 1}$.\qed

\emph{Theorem 4.} For arbitrary $N$ channels $E_1,~E_2,\cdots,E_N$, with Kraus representation $E_t(\rho)=\sum\limits_{i=1}^nK^t_i\rho(K^t_i)^\dag$, $t=1, 2,\cdots, N$, we have
\begin{equation}\label{N channel uncertain 2}
   \sum_{t=1}^NI_{\rho}(E_t)\geq\max_{\pi_t,\pi_s\in S_n}\left\{\frac{1}{N}\sum_{i=1}^nI_{\rho}\left(\sum_{t=1}^NK_{\pi_t(i)}^t\right)+\frac{2}{N^2(N-1)}\left[\sum_{i=1}^n\left(\sum_{1\leq t<s\leq N}\sqrt{I_{\rho}\left(K_{\pi_t(i)}^t-K_{\pi_s(i)}^s\right)}\right)^2\right]\right\},
\end{equation}
with equality when $N=2$, where $\pi_t\in S_n$ is an arbitrary $n$-element permutation.

\emph{Proof.}  It follows from inequality (\ref{inequality}).
 Specially, for two quantum channels $E_1$, $E_2$ with Kraus representations $E_1(\rho)=\sum\limits_{i=1}^nK^1_i\rho(K^1_i)^\dag$, $E_2(\rho)=\sum\limits_{i=1}^nK^2_i\rho(K^2_i)^\dag$, respectively, by equality $\|\textbf{a}+\textbf{b}\|^2+\|\textbf{a}-\textbf{b}\|^2=2(\|\textbf{a}\|^2+\|\textbf{b}\|^2)$, one gets
\begin{equation}\label{two channel uncertain}
 I_\rho(K^1_{\pi_1(i)})+ I_\rho(K_{\pi_2(i)}^2)=\frac{1}{2}\left[I_\rho(K^1_{\pi_1(i)}+ K_{\pi_2(i)}^2)+I_\rho(K^1_{\pi_1(i)}- K_{\pi_2(i)}^2)\right].
\end{equation}
Then \begin{equation}\label{two channel uncertain2}
 I_\rho(E_1)+ I_\rho(E_2)=\frac{1}{2}\left[\sum_{i=1}^nI_\rho(K^1_{\pi_1(i)}+ K_{\pi_2(i)}^2)+\sum_{i=1}^nI_\rho(K^1_{\pi_1(i)}- K_{\pi_2(i)}^2)\right],
\end{equation}
where  $\pi_1,\pi_2\in S_n$ are arbitrary $n$-element permutations.\qed

Clearly, our formula (\ref{two channel uncertain2}) gives a tighter uncertainty relation with larger lower bound than the UR (\ref{Luo two channel uncertain}).

\emph{Example 4.} Suppose that $\rho=\frac{1}{2}(\mathcal{I}+\vec{r}\cdot\vec{\sigma})$  with  $\vec{r}=(\frac{\sqrt{3}}{2}\cos\theta,\frac{\sqrt{3}}{2}\sin\theta,0)$, $ 0\leq\theta\leq \pi $, and three channels,  the phase damping channel $\phi$, the amplitude damping channel $\varepsilon$ and the bit flip channel $\Lambda$, where $\phi(\rho)=\sum\limits_{i=1}^2K_i\rho(K_i)^\dag$ with $K_1=|0\rangle\langle0|+\sqrt{1-q}|1\rangle\langle1|,~K_2=\sqrt{q}|1\rangle\langle1|$ and $0\leq q\leq1$,  $\varepsilon(\rho)=\sum\limits_{i=1}^2E_i\rho(E_i)^\dag$ with $E_1=|0\rangle\langle0|+\sqrt{1-q}|1\rangle\langle1|,~E_2=\sqrt{q}|0\rangle\langle1|,~0\leq q\leq1$, and $\Lambda(\rho)=\sum\limits_{i=1}^2F_i\rho(F_i)^\dag$ with operation elements $F_1=\sqrt{q}(|0\rangle\langle0|+|1\rangle\langle1|)$, $F_2=\sqrt{1-q}(|0\rangle\langle1|+|1\rangle\langle0|)$. Then according to
 \emph{Theorem 3} one has $I_{\rho}(\phi)+I_{\rho}(\varepsilon)+I_{\rho}(\Lambda)\geq \max\{A1,A2,A3,A4\}$. Here
$Ai$ $(i=1,2,3,4)$ are the lower bounds when $\{\pi_1=(1),\pi_2=(1),\pi_3=(1)\}$, $\{\pi_1=(1),\pi_2=(12),\pi_3=(12)\}$, $\{\pi_1=(1),\pi_2=(1),\pi_3=(12)\}$ or $\{\pi_1=(1),\pi_2=(12),\pi_3=(1)\}$. Similarly, one can obtain the UR of \emph{Theorem 4}, $I_{\rho}(\phi)+I_{\rho}(\varepsilon)+I_{\rho}(\Lambda)\geq \max\{B1,B2,B3,B4\}$ where $Bi$ $(i=1,2,3,4)$ are similar to $Ai$ $(i=1,2,3,4)$.

When $q=0.1$, we find that for some $\theta$ lower bound  (\ref{N channel uncertain 1}) is larger, while for some $\theta$ lower bound  (\ref{N channel uncertain 2}) is tighter, which is illustrated in Fig.5.
\begin{center}\label{Fig5}
\includegraphics[width=9cm]{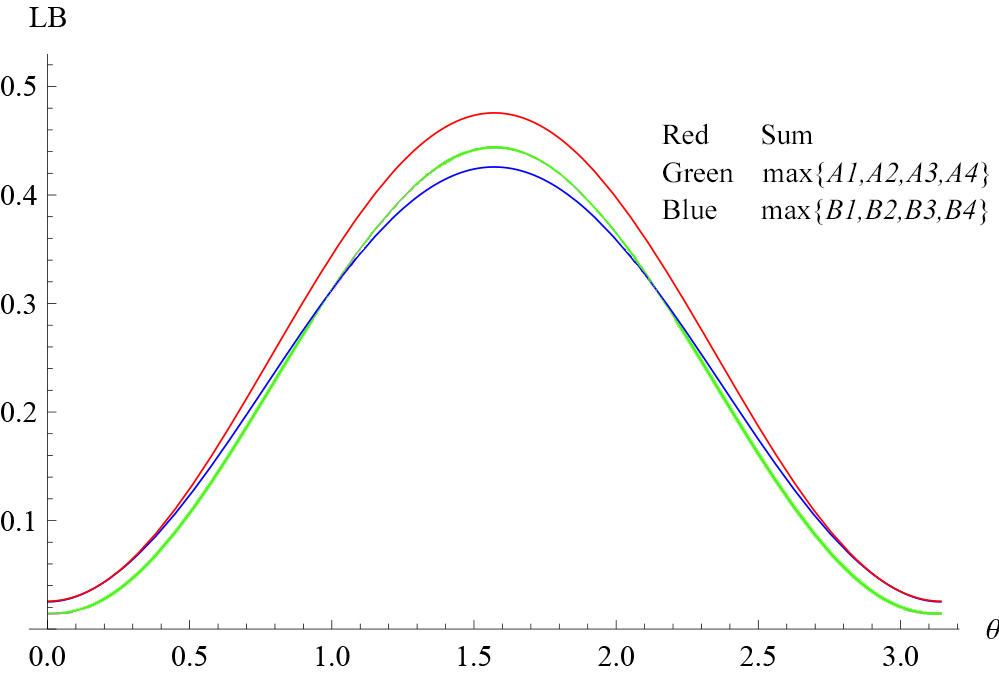}
\end{center}
{\centering\textbf{Fig.5}}~~The comparison of the lower bounds in  (\ref{N channel uncertain 1}) with lower bound in (\ref{N channel uncertain 2}) for the qubit state $\rho$ with Bloch vector $\vec{r}=(\frac{\sqrt{3}}{2}\cos\theta,\frac{\sqrt{3}}{2}\sin\theta,0)$, and three channels,  the phase damping channel $\phi$, the amplitude damping channel $\varepsilon$ and the bit flip channel $\Lambda$. The red line, the green line, and the blue line represent the sum of the skew information, the lower bound in \emph{Theorem 3}, and the lower bound in \emph{Theorem 4}, respectively.
For some cases, the UR in \emph{Theorem 3} is tighter, while sometimes it is tighter in \emph{Theorem 4}.\\
\section{Conclusion}

In summary, we obtain several sum uncertainty relations based on Wigner-Yanase skew information for finite observables and channels.
The lower bound of our uncertainty relation for $n$ incompatible observables  is nontrivial unless all observations are commutative with each other.
We provide examples in which our uncertainty relation for observables is better than all known uncertainty relations based on Wigner-Yanase skew information. We also present two uncertainty relations in terms of Wigner-Yanase skew information for multiple channels, one of which is an equality when $N=2$ and tighter than the existing UR in \cite{FuQIP2019}.  Our approach may be used to future study the uncertainty relations for multiple observables or channels.

\begin{acknowledgments}
This work was supported by the Hebei Natural Science Foundation  under Grant No. A2018205125.
\end{acknowledgments}

\end{document}